\RequirePackage{snapshot}
\documentclass[]{jfm}

\usepackage[utf8]{inputenc}
\usepackage{pgfplots}
\usepackage{graphics}
\usepgfplotslibrary{groupplots}
\usepackage{psfrag}
\usepackage{color}
\usepackage[intlimits]{amsmath}
\usepackage{amsfonts}
\usepackage{amssymb}
\usepackage{mathrsfs}
\usepackage{tikz}
\usepackage{natbib}
\usepackage{pgf}
\usetikzlibrary{arrows,cd}
\usepackage{soul}

\ifCUPmtlplainloaded \else
  \checkfont{eurm10}
  \iffontfound
    \IfFileExists{upmath.sty}
      {\typeout{^^JFound AMS Euler Roman fonts on the system,
                   using the 'upmath' package.^^J}%
       \usepackage{upmath}}
      {\typeout{^^JFound AMS Euler Roman fonts on the system, but you
                   dont seem to have the}%
       \typeout{'upmath' package installed. JFM.cls can take advantage
                 of these fonts,^^Jif you use 'upmath' package.^^J}%
      }
  \else
  \fi
\fi

\makeatletter
\newsavebox{\@brx}
\newcommand{\llangle}[1][]{\savebox{\@brx}{\(\m@th{#1\langle}\)}%
  \mathopen{\copy\@brx\kern-0.5\wd\@brx\usebox{\@brx}}}
\newcommand{\rrangle}[1][]{\savebox{\@brx}{\(\m@th{#1\rangle}\)}%
  \mathclose{\copy\@brx\kern-0.5\wd\@brx\usebox{\@brx}}}
\makeatother

\makeatletter
\newcommand{\specialnumber}[1]{%
  \def\tagform@##1{\maketag@@@{(\ignorespaces##1\unskip\@@italiccorr\emph{#1})}}%
}
\newcommand{\specialeqref}[2]{\begingroup
  \def\tagform@##1{\maketag@@@{(\ignorespaces##1\unskip\@@italiccorr\emph{#2})}}%
  \eqref{#1}\endgroup}
\makeatother

\ifCUPmtlplainloaded \else
  \checkfont{msam10}
  \iffontfound
    \IfFileExists{amssymb.sty}
      {\typeout{^^JFound AMS Symbol fonts on the system, using the
                'amssymb' package.^^J}%
       \usepackage{amssymb}%
         \let\leq=\leqslant
         \let\geq=\geqslant
      }{}
  \fi
\fi

\ifCUPmtlplainloaded \else
  \IfFileExists{amsbsy.sty}
    {\typeout{^^JFound the 'amsbsy' package on the system, using it.^^J}%
     \usepackage{amsbsy}}
    {\providecommand\boldsymbol[1]{\mbox{\boldmath $##1$}}}
\fi

\newcommand\hy{{\text -}}

\title[On the robustness of emptying filling boxes to sudden changes in the wind]{On the robustness of emptying filling boxes to sudden changes in the wind} \author[John Craske and
Graham O. Hughes]%
{John Craske$^1$\thanks{Email address for correspondence:
    john.craske07@imperial.ac.uk} and Graham O. Hughes$^1$}
\affiliation{$^1$Department of Civil and Environmental
  Engineering, Imperial College London,\\[\affilskip] London SW7 2AZ, UK}
\pubyear{}
\volume{}
\pagerange{}
\date{?; revised ?; accepted ?. - To be entered by editorial office}

\newcommand{\od}[2]{\dfrac{\mathrm{d} #1}{\mathrm{d} #2}}

\newcommand{\ez}[1]{\boldsymbol{k}}

\begin{document}

\maketitle

\begin{abstract}
  We determine the smallest instantaneous increase in {the
    strength} of an opposing wind that is necessary to permanently
  reverse the forward displacement flow that is driven by a
  two-layer thermal stratification. With an interpretation in
  terms of the flow's energetics, the results clarify why the
  ventilation of a confined space with a stably-stratified
  buoyancy field is less susceptible to being permanently reversed
  by the wind than the ventilation of a space with a uniform
  buoyancy field. For large opposing wind strengths we derive
  analytical upper and lower bounds for the system's marginal
  stability, which exhibit a good agreement with the exact
  solution, even for modest opposing wind strengths. The work
  extends a previous formulation of the problem
  \citep[][\emph{Building and Env.}  {\bf 44},
  pp. 666-673]{LisBbae2009a} by accounting for the transient
  dynamics and energetics associated with the homogenisation of
  the interior, which prove to play a significant role in
  buffering temporal variations in the wind.
\end{abstract}

\section{Introduction}
\label{sec:intro}

\subsection{Background}
\label{sec:back}

In contrast to indoor conditions that are controlled mechanically,
naturally ventilated spaces surrender themselves to the forces and
fluctuations of their surrounding environments \citep[see, for
example,][]{LinPafm1999a}. In addition to the prediction of the
steady state of a system, one should therefore be concerned with
its robustness or its propensity to switch abruptly to an
alternative steady state.

In the specific case of an indoor space subjected to a source of
heating and an external wind load, the governing equations admit
multiple steady state solutions \citep{HunGjfm2005a}. The
solutions correspond to either \emph{forward flow} or
\emph{reverse flow}, for which unidirectional discharge occurs
through the opening at the top or bottom of the space,
respectively. From an operational point of view it is necessary to
consider the transient route towards these steady states from
time-dependent governing equations \citep{KayNjfm2004a,
  CooIjfm2011a}. The analysis of the system's transient behaviour
leads naturally to questions relating to the sensitivity and
robustness of steady states to random or controlled variations in
design or environmental conditions.

There exist reverse flows that are unstable, in the sense that
infinitesimal changes in the wind's strength will cause a dramatic
change in the system's state. In contrast, there exist locally
stable reverse and forward flows that are insensitive to
infinitesimal changes in the wind. Finite changes in the wind,
however, can result in a transition between stable forward flow
and stable reverse flow, and provide the motivation for the present
study.

Previous work \citep{YuaJbae2008a, LisBbae2009a} has determined
the minimum instantaneous amount by which the strength of an
opposing wind must increase to force a transition from forward
flow to reverse flow, under the assumption of a {base} state
consisting of uniform buoyancy. Given that isolated sources of
buoyancy and heterogeneous boundary conditions result in spatially
non-uniform distributions of buoyancy \citep[see, for
example,][]{LinPjfm1990a}, we relax the assumption of uniform
buoyancy and quantify the extent to which the destruction of a
stratified interior modifies the system's robustness to
fluctuations in the wind.

\begin{figure}
  \small
  \begin{center}
  \includegraphics{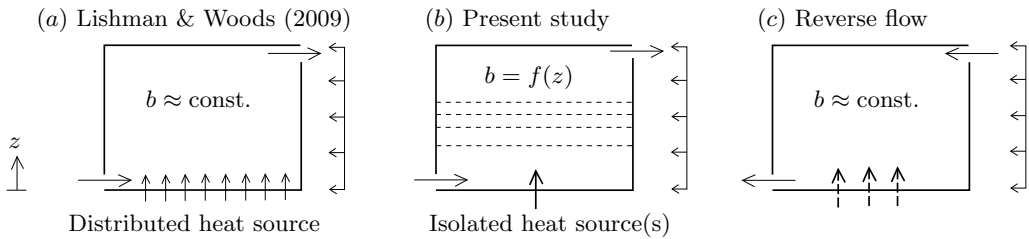}
\end{center}
\caption{The ventilation driven by $(a)$ a well-mixed interior of
  uniform buoyancy and $(b)$ a stably-stratified
  interior. {Schematic} $(c)$ illustrates reverse flow, for
  which the interior is well mixed and assumed to be of uniform
  buoyancy.}
  \label{diag:05}
\end{figure}

\subsection{An illustration of the general problem}
\label{sec:example}

Consider a volume with low- and high-level openings, as depicted
in figure \ref{diag:05}. In $(a)$, following \citet{LisBbae2009a},
the heating of the space is distributed evenly over the floor and
the resulting buoyancy field is assumed to be uniform. In $(b)$,
which is the starting point for the present work, the heating
occurs unevenly over localised sources to produce a buoyancy field
that is non-uniform (i.e. stratified). Ventilation of the space is
driven by pressure differences resulting from the average internal
buoyancy and external forces arising from the wind. In a steady
state, the rate at which buoyancy drains from the top of the space
is equal to the rate at which it is supplied to the space in the
form of heat, which we assume to be the same in figure \ref{diag:05}$(a)$-$(c)$.

The average buoyancy and ventilation rate is greater in figure
\ref{diag:05}$(a)$ than in figure \ref{diag:05}$(b)$ \citep[see, for
example,][\S 4.3]{GlaCjfm2001a}. To understand why, note that the
ventilation rate increases with pressure difference across the
upper opening, which in turn increases with average buoyancy in
the volume. For a given average buoyancy, the buoyancy at the top
of the volume in figure \ref{diag:05}$(a)$ is less than it is for any stable
stratification figure \ref{diag:05}$(b)$; hence the buoyancy flux through the upper
opening will only be the same in both cases if the ventilation
rate -- and therefore the average buoyancy -- is maximised in
figure \ref{diag:05}$(a)$.

Figure \ref{diag:05}$(c)$ illustrates a situation in which the
pressure difference across the space due to wind exceeds the
pressure difference created by internal buoyancy. The resulting
flow is in the reverse direction and is accompanied by an
approximately well-mixed interior of uniform buoyancy, regardless
of the way in which the space is heated \citep{HunGjfm2005a}.  In
a steady-state, the average buoyancy in $(c)$ is necessarily less
than the average buoyancy in figure \ref{diag:05}$(b)$ and,
therefore, less than the average buoyancy in figure
\ref{diag:05}$(a)$.

A transient increase in the opposing wind strength can cause a
transition from forward flow to reverse flow. The question that
our work addresses is whether the minimum increase in the wind
strength that is required for the transition is greater for the
system in figure \ref{diag:05}$(a)$ than it is for the system in
figure \ref{diag:05}$(b)$. Whilst the required reduction in
average buoyancy is greater for the system in figure
\ref{diag:05}$(a)$, we show that the wind must perform additional
work over a finite time to destroy the stratification for the
system in figure \ref{diag:05}$(b)$, making it more robust to
fluctuations in the wind than one might otherwise expect.

\subsection{Theoretical model}
\label{sec:govern}

\begin{figure}
  \begin{center}
  \includegraphics{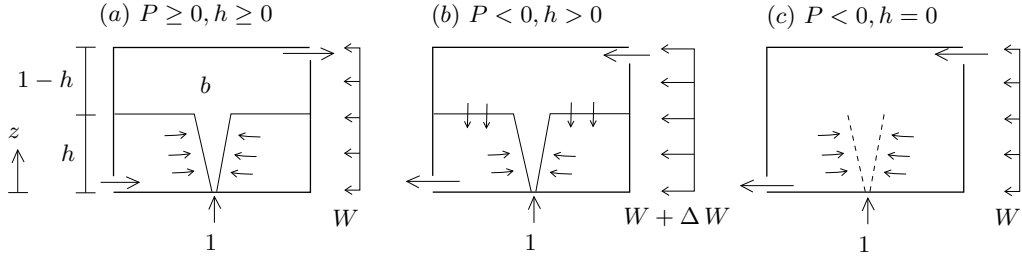}
\end{center}
\caption{Wind-opposed buoyancy driven ventilation resulting in
  $(a)$: steady forward flow; $(b)$: transient reverse flow with a
  stratification and $(c)$: steady reverse flow. A sufficiently large
  increase $\Delta W$ in the opposing wind $W$ results in a transition from $(a)$
  to $(c)$.}
  \label{diag:04}
\end{figure}

We will assume that buoyancy is introduced from a point source
located at the bottom of a domain, whose ventilation is
facilitated by low- and high-level openings. For forward flow, the
resulting two-layer stratification, illustrated in figure
\ref{diag:04}$(a)$, is a special example of the stratified
environments that were considered in figure
\ref{diag:05}$(b)$. The state of the system can be described by
the dimensionless height of the resulting interface $h$ and the
dimensionless {uniform buoyancy $b$ of a well-mixed upper layer},
which, following \citet{CooIjfm2011a}, are non-dimensionalised
using the total domain height and plume buoyancy flux. Following
\citet{HunGjfm2005a}, the volume flow rate through the space is
determined by the dimensionless pressure difference
\begin{equation}
  P = \underbrace{b(1-h)}_{\text{buoyancy}}-\underbrace{W}_{\text{wind}},
\label{eq:Q}
\end{equation}
\noindent between the dimensionless stack-induced pressure
difference $b(1-h)$ and the dimensionless wind-induced pressure
difference $W$ between the windward and leeward openings, where
$W$ corresponds to the square of a Froude number {based on the wind
  speed}. In adopting this notation, which differs slightly from
the explicit use of the Froude number $Fr\propto\sqrt{W}$ by
\citet{HunGjfm2005a} to express the same physical concepts, we
follow \citet{CooIjfm2011a}, in which further details pertaining
to the non-dimensionalisation can be found.

In an unsteady state the governing equations comprise statements
of volume conservation and buoyancy conservation in the upper layer:
\begin{equation}
\od{h}{t} =
\begin{cases}
  -h^{5/3}+|VP|^{1/2},\\
  -h^{5/3}-|VP|^{1/2},\\
  0,
\end{cases}
\od{}{t}b(1-h) =
\begin{cases}
  1-|VP|^{1/2}b,\ & P\geq 0, h\geq 0\\
  1,\ &  P<0, h>0\\
  1-|VP|^{1/2}b,\ & P<0, h=0
\end{cases}
\label{eq:h}
\specialnumber{a,b}
\end{equation}
\noindent respectively, where $V$ is a dimensionless opening area
that accounts for discharge coefficients. The $h^{5/3}$ in
\specialeqref{eq:h}{a} corresponds to the volume flux in an {axisymmetric
  plume} at $z=h$, which would cause the height of the interface
to reduce in the absence of the stack-driven discharge
$|VP|^{1/2}$. The sub-equations in \eqref{eq:h} 
refer to forward displacement ventilation ($P\geq 0$, $h\geq 0$),
reverse displacement ventilation ($P<0$, $h>0$) and reverse mixing
ventilation ($P<0$, $h=0$), as depicted in figure
\ref{diag:04}. If a well-mixed interior of uniform buoyancy is
assumed from the outset \citep[as it is in][]{LisBbae2009a}, the
system's state can be uniquely described by the buoyancy $b$
alone, which evolves according to \specialeqref{eq:h}{b} with $h=0$.

When $W$ exceeds a critical value $W_{c}=\sqrt[3]{27/4V}$, the
system has three fixed points, at which
$\mathrm{d}(b,h)/\mathrm{d}t=0$, each corresponding to a different
steady-state solution \citep{HunGjfm2005a}. For $W>W_{c}$ there
are two stable solutions and an unstable steady-state solution
describing reverse flow. Reverse flow is not possible when
$W<W_{c}$ and, as pointed out by \citet{LisBbae2009a}, {stable
  reverse flow} subjected to decreasing wind will jump to
displacement ventilation when $W<W_{c}$.

The fixed point for forward flow $(b_{0}, h_{0})$ through a
stratified environment satisfies \eqref{eq:h} for $P\geq 0$:
\begin{equation}
  b_{0}=h_{0}^{-5/3},\quad\quad Vh_{0}^{-5/3}(1-h_{0})-h_{0}^{10/3}-VW=0.
  \label{eq:fpf}
\end{equation}

\begin{figure}
  \begin{center}
  \includegraphics{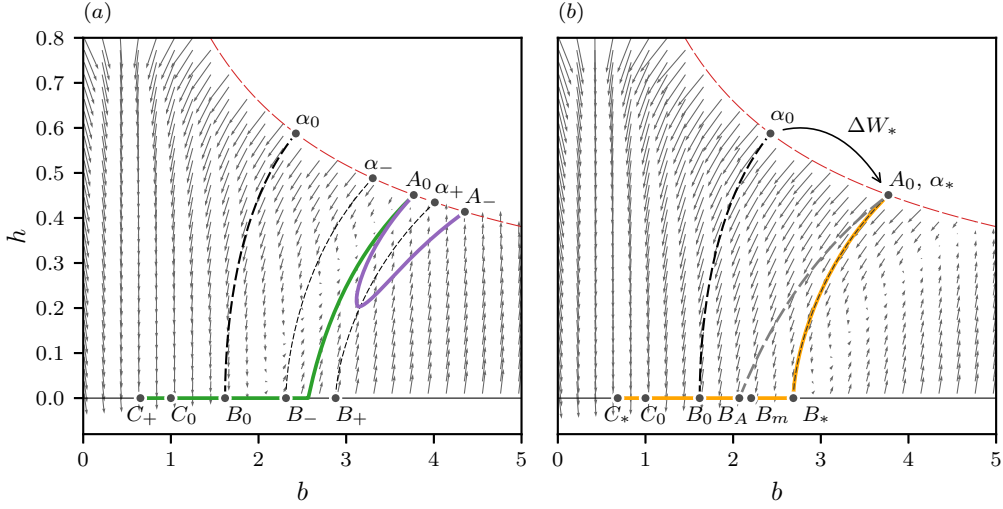}
\end{center}
\caption{The {projected} phase space for a two-layer
  stratification {with base wind strength $W=2$} and $V=1$.  The
  points $A_{0}$, $B_{0}$ and $C_{0}$ denote the stable (forward
  flow), unstable and stable (reverse flow) fixed points of the
  system, respectively. The line $\alpha_{0}\hy B_{0}$ denotes the
  separatrix curve for $W=2$ and $V=1$. The lines $A_{0}\hy A_{-}$
  and $A_{0}\hy C_{+}$ in $(a)$ denote the trajectories taken by
  the system following sub-optimal $\Delta W = 0.5<\Delta W_{*}$
  and super-optimal $\Delta W = 1.0>\Delta W_{*}$ step changes in
  the wind strength, respectively. The thin dashed lines
  $\alpha_{-}\hy B_{-}$ and $\alpha_{+}\hy B_{+}$ denote the
  position of the separatrix curves for sub- and super-optimal
  changes in the wind, respectively. The line $A_{0}\hy C_{*}$ in
  $(b)$ denotes the trajectory taken by the system following an
  optimal $\Delta W_{*}\approx 0.825$ step change in the wind
  strength, which, for $h>0$ coincides with the system's
  separatrix curve $\alpha_{*}\hy B_{*}$ for the wind strength
  $W + \Delta W_{*}$. Along $A_{0}\hy B_{A}$ the average buoyancy
  $b_{0}(1-h_{0})$ is constant and $B_{A}$ corresponds to the
  location of the unstable fixed point for a sub-optimal step
  change $\Delta W<\Delta W_{*}$ in wind strength. The point
  $B_{m}$ is the location of the fixed point for forward flow,
  under the assumption of a well-mixed interior
  \citep{LisBbae2009a}. {The grey arrows denote the time
    derivatives of a given state $(b, h)$ for an opposing wind of
    strength $W$ in $(a)$ and $W+\Delta W_{*}$ in $(b)$.}}
  \label{fig:14}
\end{figure}

\noindent A fixed point for reverse flow when $W>W_{c}$ satisfies
\specialeqref{eq:h}{b} for $h=0$ and is therefore a stable or
unstable state corresponding to one of two positive real roots of
the cubic
\begin{equation}
  b^{3}-Wb^{2}+ V^{-1}=0.
  \label{eq:fpr}
\end{equation}
{It is useful to regard the two-dimensional phase space for the
  system, shown in figure} \ref{fig:14}, {as a projection of the
  space to which states $(b,h,W,V)$ belong. Constant values of $V$
  and $W$ correspond to a particular plane or slice through the
  entire space. The features of phase space that are shown in the
  projection in figure \ref{fig:14} therefore depend on particular
  values of $V$ and $W$, whose axes are hidden from view. {Whilst
    the grey arrows in figure} \ref{fig:14}$(a)$ correspond to the
  system's time derivatives when $V=1$ and $W=2>W_{c}$, we have
  also included the system's trajectory for wind strengths
  $W+\Delta W$ to indicate how the system would evolve if the base wind
  strength $W$ were to change by $\Delta W$.}

As discussed in \citet{CooIjfm2011a}, states for which
$h>b^{-3/5}$, representing an upper layer whose buoyancy exceeds
the buoyancy in the plume at the interface, are beyond the scope
of the model equations and therefore not included in figure
\ref{fig:14}. For a given base wind strength $W\geq W_{c}$, the
phase space is partitioned by a separatrix curve
$B_{0}\hy \alpha_{0}$, which emanates from the unstable fixed
point $B_{0}$, into basins of attraction corresponding to the
stable fixed points $A_{0}$ and $C_{0}$. The fixed point to which
a system's state eventually evolves is determined by whether its
state lies to the left or to the right of the separatrix
curve. {The separatrix curve can be obtained by adding a small
  positive perturbation to $h=0$ at the unstable fixed point
  $B_{0}$, to provide initial conditions for the integration of
  the governing equations \emph{backwards} in time until
  $h=b^{-3/5}$ at $\alpha_{0}$.}
\section{Flow reversal}
\label{sec:results}

The direction of the flow through the system can be permanently
reversed by a sustained increase $\Delta W$ in the wind
strength. The resulting pressure difference must be larger than
the favourable pressure difference created by the upper layer of
warm fluid during the transition. 

\subsection{The critical change $\Delta W_{*}$ in the wind strength}

If the system's state starts at the fixed point $A_{0}$
corresponding to steady forward flow for a base wind strength $W$,
a step increase $\Delta W$ in the wind strength will cause the
state to change. Hereafter we refer to $\Delta W_{*}$ as the
minimum increase required to permanently reverse the flow.  Thus,
when $\Delta W < \Delta W_{*}$, the system returns to a state
corresponding to forward flow (see, for example, the line
$A_{0}\hy A_{-}$ in figure \ref{fig:14}$(a)$, projected from the
plane $W+\Delta W$). When $\Delta W > \Delta W_{*}$ the system
transitions to a state corresponding to stable reverse flow (see,
for example, the line $A_{0}\hy C_{+}$ in figure
\ref{fig:14}$(a)$, projected from the plane $W+\Delta W$). Whether
a transition to stable reverse flow occurs depends on whether the
step increase $\Delta W$ moves the separatrix curve to the left or
to the right of the system's {base} state $A_{0}$.


The strength of the optimal (minimum) wind {increase}
$\Delta W_{*}:\alpha_{0}\mapsto\alpha_{*}$ {projects} the point
$\alpha_{*}$ of the separatrix curve in the $W+\Delta W_{*}$ plane
exactly onto the fixed point $A_{0}$ for forward flow
corresponding to $W$, as shown in figure \ref{fig:14}$(b)$. The
{increase} in $W$ would need to be {sustained} for at least as long as it
would take for the system's trajectory to cross the separatrix
curve at the unstable fixed point $B_{0}$ associated with the
{base} wind strength $W$. No instantaneous {increase} for which
$\Delta W < \Delta W_{*}$ can reverse the flow because
$\Delta W_{*}$ is the smallest step change that places the
system's state in the basin of attraction for stable reverse flow
{in the $W$ plane}.

The base wind strength that corresponds to $W+\Delta W_{*}$ can be
found by integrating the governing equations backwards in time
along the separatrix curve $B_{*}\hy \alpha_{*}$ from $(b,h)=(b_{*},0)$
for $W+\Delta W_{*}$, {where $b_{*}$ satisfies a modified version of} \eqref{eq:fpr}:
\begin{equation}
  b_{*}^{3}-(W+\Delta W_{*})b_{*}^{2}+V^{-1}=0.
  \label{eq:fprWc}
\end{equation}
\noindent The point $\alpha_{*}$, at which the resulting trajectory
intersects the line $h=b^{-3/5}$, corresponds to the steady-state
solution $A_{0}$ for forward flow for a base wind {strength}
$W$. Performing the calculation for different values of
$W+\Delta W_{*}$ provides the relationship between $W$ and
$\Delta W_{*}$ for marginal stability that is displayed in figure
\ref{fig:13}.

\subsection{Comparison with \citet{LisBbae2009a}}

\begin{figure}
  \begin{center}
  \includegraphics{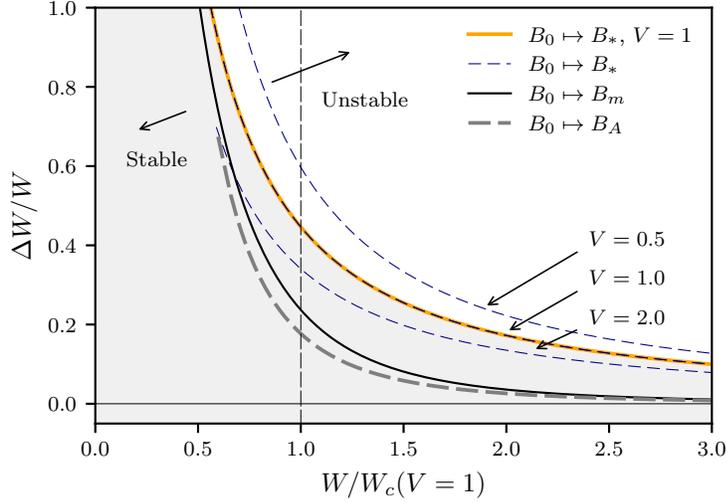}
\end{center}
\caption{{Stability diagram indicating instantaneous increases in
    the wind strength $\Delta W$ that cause a reversal of forward
    ventilation flow}. The stable (shaded) and unstable (unshaded)
  regions are separated by the minimal instantaneous increase in
  wind strength $\Delta W_{*}: B_{0}\mapsto B_{*}$ that is
  necessary to permanently reverse the ventilation of a two-layer
  stratified interior for an opening parameter $V=1$ (the dashed
  lines corresponding to $V=0.5$ and $V=2.0$ are included for
  comparison). The lines in the stable region correspond to the
  unstratified interior of uniform buoyancy considered by
  \citet{LisBbae2009a}, corresponding to $B_{m}$ in figure
  \ref{fig:14}$(b)$ and the wind increase estimated from the
  average buoyancy $b_{0}(1-h_{0})$ of a two-layer stratification,
  corresponding to $B_{A}$ in figure \ref{fig:14}$(b)$.}
\label{fig:13}
\end{figure}

\noindent A steady-state forward flow in the environment of
uniform buoyancy considered by \citet{LisBbae2009a} satisfies
\specialeqref{eq:h}{b} with $P\geq 0$ and $h=0$:
\begin{equation}
  b_{m}^{3}-Wb_{m}^{2}-V^{-1}=0.
  \label{eq:fpm}
\end{equation}
\noindent whose real solution corresponds to $B_{m}$ in figure
\ref{fig:14}$(b)$. It is evident from figure \ref{fig:14}$(b)$
that $b_{m}<b_{*}$, where $b_{*}$ is the buoyancy of the
stratified interior's {`upper'} layer {when}, during application
of the step increase in the wind strength, the interface reaches
floor level and the upper layer engulfs the entire space. The
minimum increase in the wind necessary to reverse the flow through
an initially stratified environment is therefore greater than it
is for an environment whose buoyancy is initially uniform. As can
be seen in figure \ref{fig:13}, for base wind strengths close to
$W_{c}=\sqrt[3]{27/4V}$, the critical {increase} $\Delta W_{*}$
for a two-layer stratification is approximately twice as large as
it is for an interior of uniform buoyancy, for which
$\Delta W_{*}=2(b_{m}-W)$ \citep{LisBbae2009a}. At larger base
wind strengths $W\gg W_{c}$ a stratified interior can withstand
changes in the wind that are an order of magnitude larger than
those that can be withstood by an interior of uniform buoyancy.

We have demonstrated that states of uniform buoyancy satisfying
{\eqref{eq:fpm}} are less robust to changes in the wind than
states consisting of a two-layer stratification satisfying
{\eqref{eq:fpf}}. As explained in {\S\ref{sec:example}}, given
that the average buoyancy of the two-layer stratification is less
than the uniform buoyancy satisfying {\eqref{eq:fpm}}, we expect
estimations of the system's robustness based on its average
buoyancy $b_{0}(1-h_{0})$ to be misleading and
conservative. Indeed, as deduced in {\S\ref{sec:example}}, the
line $A_{0}\hy B_{A}$ in figure \ref{fig:14}$(b)$, on which
$b(1-h)=b_{0}(1-h_{0})$ is constant, intersects the {horizontal
  axis} $h=0$ at $B_{A}$, where
$b=b_{0}(1-h_{0})<b_{m}$. Conclusions about the robustness of
stratified environments based on their average buoyancy therefore
provide a lower bound on the strength of the minimum destabilising
increase in the wind, which suggests that the work required to
homogenise a stratified interior plays a crucial role in
determining its stability.

  Changing the opening parameter $V$ affects neither the
  qualitative aspects of our results concerning the system's
  robustness nor our comparisons with interiors of uniform
  buoyancy. As shown in figure \ref{fig:13}, and consistent with
  intuition, large values of the opening parameter, corresponding
  to openings with relatively large area, make the stratified
  interior more susceptible to flow reversal, whilst small values
  make it more robust.  It is interesting that the enhanced
  robustness afforded by a reduction in $V$ also entails a
  reduction in the height of the interface of a two-layer
  stratification.

  \subsection{Time scale}
  \label{sec:timescales}

  The transient route that forward displacement flow takes before
  being permanently reversed by a sufficiently large increase
  $\Delta W_{*}$ in the wind strength comprises two distinct
  processes. The first involves lowering the interface of the
  stratification until the warm upper layer engulfs the entire
  space ($A_{0}\hy B_{*}$ in figure \ref{fig:14}). The second
  involves purging warm air from the space via the lower opening,
  until the system reaches stable equilibrium ($B_{*}\hy C_{*}$ in
  figure \ref{fig:14}). The subject of this section is the time
  scale on which these processes occur.

\begin{figure}
  \begin{center}
 \includegraphics{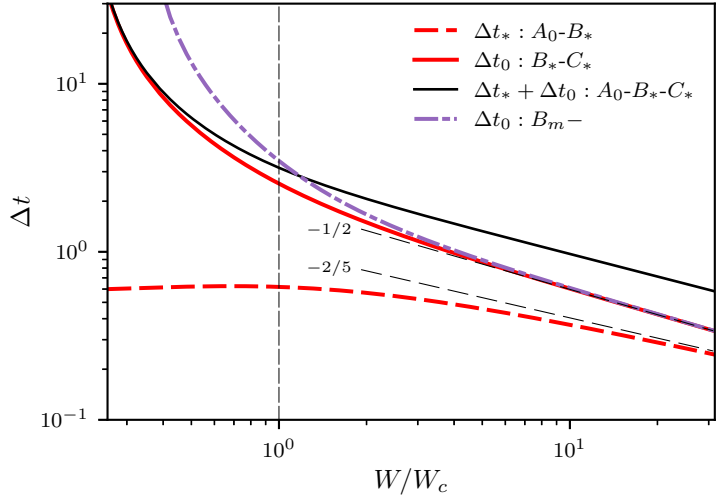}
\end{center}
\caption{Transient time scale associated with lowering the
  interface of a stratified interior along $A_{0}\hy B_{*}$ in
  figure \ref{fig:14} ($\Delta t_{*}$, solved numerically) and
  purging relatively warm well-mixed air from the space along
  $B_{*}\hy C_{*}$ ($\Delta t_{0}$, estimated using \eqref{eq:tB})
  for an opening parameter $V=1$. The curve labelled `$B_{m}\hy$'
  corresponds the timescale estimated from \eqref{eq:tB} for a
  well-mixed state to move from $B_{m}$ to stable reverse
  flow. The thin dashed lines, whose gradients are $-1/2$ and
  $-2/5$ correspond to the asymptotic scaling of $\Delta t_{*}$
  and $\Delta t_{0}$ when $W\rightarrow\infty$, as discussed in
  \S\ref{sec:bounds}.}
  \label{fig:17}
\end{figure}

  The time $\Delta t_{*}$ that it takes for the interface to be
  lowered to floor level can be obtained by numerical integration,
  backwards in time, along $B_{*}\hy A_{0}$, and is displayed in
  figure \ref{fig:17}. A consequence of the discontinuity at $h=0$
  in the governing equations \eqref{eq:h} is that the time
  derivatives of $b$ and $h$ are non-zero arbitrarily close to the
  unstable fixed point $B_{*}$. Physically, this corresponds to
  the fact that the interface descends with a finite speed just
  before it reaches floor level.
  
  In contrast to the trajectory $A_{0}\hy B_{*}$, temporal
  derivatives tend to zero at both ends of the trajectory
  $B_{*}\hy C_{*}$, which makes the time scale associated with the
  purging of warm air from the space infinite without a
  perturbation at $B_{*}$. Using \eqref{eq:h} when $P<0,\ h=0$,
  for the modified wind strength $W+\Delta W_{*}$, we define a
  characteristic timescale by noting that
  $-\mathrm{d}b/\mathrm{d}t$ is maximised when
  $b=2(W+\Delta W_{*})/3$; hence

  \begin{equation}
    -\od{b}{t} \leq \left( \frac{W+\Delta W_{*}}{W_{c}} \right)^{3/2}-1,
  \end{equation}

  \noindent along $B_{*}\hy C_{*}$. A useful lower bound on the
  time $\Delta t_{0}$ that it takes for the system to travel along
  $B_{*}\hy C_{*}$ is therefore

\begin{equation}
  \frac{\Delta b}{\left( \dfrac{W+\Delta W_{*}}{W_{c}} \right)^{3/2}-1}\leq\Delta t_{0}.
  \label{eq:tB}
\end{equation}
  
\noindent where $\Delta b\geq 0$ is the difference in buoyancy between
the two fixed points for reverse flow from \eqref{eq:fpr}. The
time scale $\Delta t_{0}$ is displayed in figure \ref{fig:17} for $V=1$,
along with the total time $\Delta t_{*}+\Delta t_{0}$. Note that
the estimation \eqref{eq:tB} relates to the time taken for the
system to reach $C_{*}$, and does not, therefore, necessarily
provide an indication of the minimum duration for which the
increase in wind strength $\Delta W_{*}$ would be need to be
sustained. The latter would involve estimating the time it would
take for the system to travel along $B_{*}\hy B_{0}$.

The lower bound \eqref{eq:tB} indicates that the time
$\Delta t_{0}$ that it takes for the system to travel along
$B_{*}\hy C_{*}$ is substantially larger than the time
$\Delta t_{*}$ that it takes for the system to travel along
$A_{0}\hy B_{*}$, for the base wind speeds shown in figure
\ref{fig:17}. Both $\Delta t_{*}$ and $\Delta t_{0}$ decrease with
increasing base wind strength $W$. Consequently, the estimation
\eqref{eq:tB} for a well-mixed interior, for which $\Delta W_{*}$
is less than it is for a stratified interior, is always greater
than $\Delta t_{0}$ for a stratified interior, although the
difference is insignificant for $W/W_{c}\gtrapprox 2$.

To interpret figure \ref{fig:17} from a practical perspective it
is useful to note that time in \eqref{eq:h} is non-dimensionalised
using the `filling-box' time scale \citep{CooIjfm2011a}
$S/(cF^{1/3}H^{2/3})$, where $S$ is the horizontal area of the
space, $c=6\varepsilon/5(9\varepsilon\pi^{2}/10)^{1/3}$ for an
entrainment coefficient $\varepsilon$, $F$ is the buoyancy flux
and $H$ is the height of the space. For reference, if
$S=100\,\mathrm{m}^{2}$, $F=0.1\,\mathrm{m}^{4}\mathrm{s}^{-3}$
(corresponding to a heat load of approximately
$3.5\,\mathrm{kW}$), $H=5\,\mathrm{m}$ and $\varepsilon=0.1$, the
filling box time is approximately ten minutes. With this
information in mind, $\Delta W_{*}$ can only be regarded as a
`gust' in the wind for sufficiently small filling box time scales
or sufficiently large base wind speeds. It is otherwise more
appropriate to regard $\Delta W_{*}$ as arising from a more
persistent change in prevailing wind conditions.

\section{Bounds on the stability of forward flow}
\label{sec:bounds}
\subsection{Exact bounds}

{To understand the physics and scaling laws behind the
  stability of forward flow, we consider the trajectory that the
  system takes through phase space when it is subjected to an
  optimal increase in wind strength $\Delta W_{*}$.} Rather than
{retaining} the volume flux $h^{5/3}$ in
\specialeqref{eq:h}{a}, which depends on the unknown and variable
interface height $h$, we assume the volume flux to be equal to the
constant $\lambda h_{0}^{5/3}$, where $h_{0}$ is the initial
interface height. Incorporation of the volume flux in this way
results in a tractable differential equation and facilitates the
over and under estimation of the effect that the plume's volume
flux has on the system's stability. By substituting the solution
$b(1-h)=b_{*}+t$ of \specialeqref{eq:h}{b} into {the modified}
\specialeqref{eq:h}{a} for $P<0, h>0$ using the buoyancy $b_{*}$ at the unstable
fixed point, one obtains

\begin{equation}
  \od{h}{t}=-\lambda h_{0}^{5/3}
-V^{1/2}\left(W+ \Delta W_{*}-b_{*}-t\right)^{1/2}.
  \label{eq:dhdt_bnd}
\end{equation}

\noindent Along the separatrix curve, $h$ varies monotonically
between $h_{0}$ and $0$, which means that \eqref{eq:dhdt_bnd} with
$\lambda\in [0,1]$ bounds the actual solution {to} \specialeqref{eq:h}{a}. Noting that the
system is to be integrated backwards in time from $(b_{*},0)$,
$\lambda =0$ corresponds to neglecting the volume flux in the
plume, underestimating the {ascent rate of the interface} and
overestimating the optimal change in wind strength $\Delta
W_{*}$. Conversely, $\lambda =1$ corresponds to overestimating the
volume flux in the plume, overestimating the {ascent rate of
  the interface} and underestimating the optimal change in wind
strength $\Delta W_{*}$. In either case, \eqref{eq:dhdt_bnd}
provides a good approximation for the small interface heights that
occur for large $W$, because they entail a {relatively} small volume
flux term.

\begin{figure}
  \begin{center}
  \includegraphics{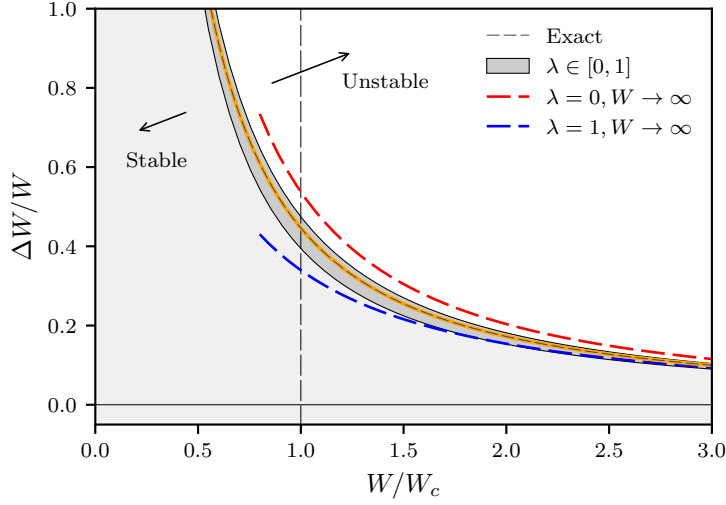}
\end{center}
\caption{Upper and lower bounds of the optimal wind increase
  $\Delta W_{*}$, indicated by the dark grey region in the
  vicinity of the marginal stability curve (thin dashed line). The
  long dashed lines $\lambda=0$ and $\lambda=1$ correspond to
  asymptotic expressions for the upper and lower bounds,
  respectively, from \eqref{eq:dWa}.}
  \label{fig:16}
\end{figure}

Unlike the original differential equation,
\eqref{eq:dhdt_bnd} is readily solved analytically to give

\begin{equation}
    h_{0}=\lambda\,h_{0}^{5/3}\Delta t_{*}+\frac{2V^{1/2}}{3}\left(W+ \Delta W_{*} -b_{*} +t\right)^{3/2}\big |_{0}^{\Delta t_{*}},
  \label{eq:h_bnd}
\end{equation}

\noindent where $\Delta t_{*}=b_{*}-b_{0}(1-h_{0})$ is the time it
takes the system to reach the unstable fixed point. Equation
\eqref{eq:h_bnd} provides the algebraic bounds of the shaded
region displayed in figure \ref{fig:16}, which exhibits a close
agreement with the exact curve for the system's marginal
stability, illustrating that the volume flux does not play a
significant role in determining the system's stability.

\subsection{Asymptotic scaling for $W\rightarrow \infty$}

For forward flow \eqref{eq:fpf} implies that the buoyancy and
interface height at the stable fixed point are, respectively:
\begin{equation}
    b_{0} = W+W^{2/5}+O\left(\frac{1}{W^{1/5}}\right),\quad h_{0}=\frac{1}{W^{3/5}}-\frac{3}{5}\frac{1}{W^{6/5}}+O\left(\frac{1}{W^{9/5}}\right),
\label{eq:bfhf}
  \end{equation}
\noindent {for fixed $V$}, which implies that the total
buoyancy of the upper layer at the stable fixed point scales
according to

\begin{equation}
  b_{0}(1-h_{0})= W + \frac{1}{VW^{2}}+O\left(\frac{1}{W^{13/5}}\right).
  \label{eq:buoy}
\end{equation}

\noindent As $W\rightarrow\infty$ the stable interface height
$h_{0}\rightarrow 0$ and \eqref{eq:fpm}, describing the
well-mixed interior assumed by \citet{LisBbae2009a}, indicates
that, to leading order, the buoyancy of the stable fixed point for
forward flow in a well mixed interior scales in the same way as
$b_{0}(1-h_{0})$:

\begin{equation}
  b_{m}= W + \frac{1}{VW^{2}}+O\left(\frac{1}{W^{5}}\right).
  \label{eq:buoy2}
\end{equation}

From \eqref{eq:fprWc}, the buoyancy at the unstable fixed point
$B_{*}$ for reverse flow at wind strength $W+\Delta W_{*}$,
assuming that $\Delta W_{*}=o(W)$, is:

\begin{equation}
    b_{*} \sim W+\Delta W_{*}-\frac{1}{VW^{2}}.
  \label{eq:gr}
\end{equation}

\begin{figure}
\begin{center}
  \includegraphics[trim=19cm 21.5cm 20cm 4cm, clip]{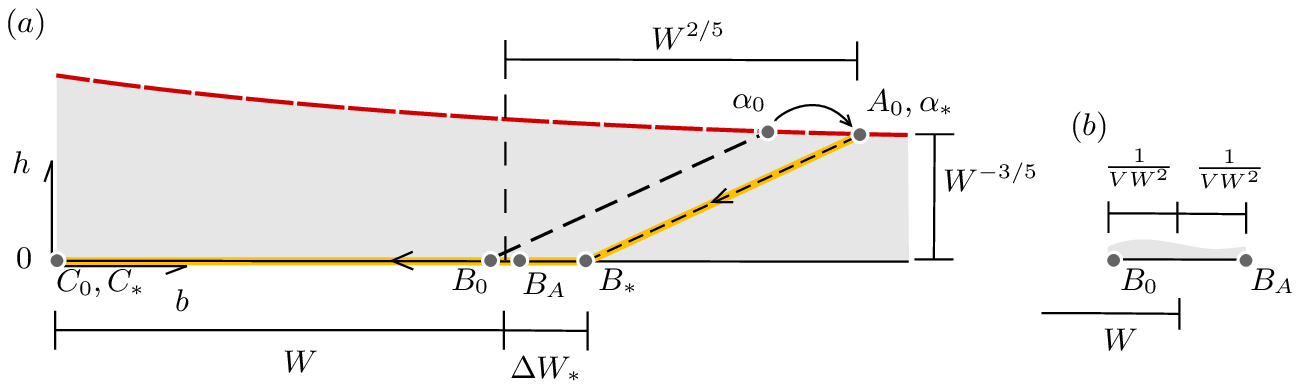}
\end{center}

\caption{The asymptotic representation of phase space (cf. figure
  \ref{fig:14}$(b)$) for $W\rightarrow \infty$, where $(a)$
  depicts the region traversed by the system's trajectory and
  $(b)$ depicts the relatively small distance between the original
  unstable fixed point $B_{0}$ (obtained from \eqref{eq:gr} with
  $\Delta W_{*}=0$) and the average buoyancy $B_{A}$.}
  \label{diag:02}
\end{figure}

\noindent Figure \ref{diag:02} illustrates the scaling of the
features of phase space when $W\rightarrow\infty$. The interface
height is small in comparison with the distance between the stable
fixed points for forward and reverse flow. {Note that
  $b\sim 1/\sqrt{VW}\rightarrow 0$ is a solution to}
  \eqref{eq:fpr} {for $W\rightarrow \infty$ and corresponds to a
  stable reverse flow in which the interior is rapidly flushed by
  the wind.} In order to destabilise forward flow, the buoyancy
associated with the unstable fixed point must increase by at least
$\Delta W_{*}$, whose scaling with respect to $W$ we determine below.
Recalling that $\Delta t_{*}=b_{*}-b_{0}(1-h_{0})$ from
buoyancy conservation, \eqref{eq:buoy} and \eqref{eq:gr}
imply that $\Delta t_{*}\sim \Delta W_{*}$, which, on substitution into
\eqref{eq:h_bnd} with \eqref{eq:bfhf}, gives
\begin{equation}
  \frac{2V^{1/2}}{3}\Delta W_{*}^{3/2}+\lambda\frac{\Delta W_{*}}{W}-\frac{1}{W^{3/5}}\sim 0, 
\label{eq:dWa}
\end{equation}
\noindent which is a cubic in $\sqrt{\Delta W_{*}}$ for any given
$\lambda\in[0,1]$. As indicated in figure \ref{fig:16},
\eqref{eq:dWa} provides a close agreement with the exact bounds
obtained in the previous section, even for relatively small values
of $W$. In particular, when $\lambda=0$, \eqref{eq:dWa}
provides the useful upper bound:
\begin{equation}
\Delta W_{*}\sim \left(\frac{3}{2}\right)^{2/3}\frac{1}{V^{1/3}W^{2/5}}.
\label{eq:dWa0}
\end{equation}
\noindent More generally, all solutions to \eqref{eq:dWa} scale
according to $\Delta W_{*}=O(1/W^{2/5})$, which is significantly
weaker than the scaling $\Delta W_{*}\sim 1/VW^{2}$ that is implied
by \eqref{eq:buoy}-\eqref{eq:gr}, based on the
assumption of a well-mixed interior driven by distributed heating
\citep{LisBbae2009a}, or a conserved average buoyancy in the case
of localised heating. 

The asymptotic time scale $\Delta t_{*}\sim \Delta W_{*}$ using
\eqref{eq:dWa0} is included in figure \ref{fig:17} alongside the
lower-bound scaling $\Delta t_{0}\sim \sqrt{W_{c}^{3}/W}$ using
$\Delta b\sim W$ in \eqref{eq:tB}.

\section{Energetics for $W\rightarrow\infty$}
\label{sec:energy}

We will now demonstrate that the scaling \eqref{eq:dWa0} accounts
for the work performed by the wind to homogenise the interior. The
optimal increase in wind strength $\Delta W_{*}$ instantaneously
reverses the flow. Therefore, over the finite time it takes for
the interface to be lowered from $h_{0}\ll 1$ to zero, the change in potential
energy $\Delta E_{p}$ is:
\begin{equation}
  \Delta E_{p}=\frac{b_{0}(1-h_{0})h_{0}}{2}-\frac{\Delta t_{*}}{2}\sim \frac{W^{2/5}}{2}.
\label{eq:Ep}
\end{equation}
\noindent The first term on the right-hand side of \eqref{eq:Ep}
accounts for lowering the centre of mass of the average buoyancy
$b_{0}(1-h_{0})$ by $h_{0}/2$, while the second term accounts for
buoyancy added (uniformly) to the domain over
$\Delta t_{*}\sim\Delta W_{*}\propto 1/W^{2/5}$ {at an average
  height of $1/2$}, which is comparatively small.

Neglecting the $O(\Delta t_{*}h_{0})$ work performed by buoyancy
in the lower layer, the change in the system's potential energy is
equal to a fraction $\eta\in [0,1]$ of the total work
$\Delta E_{w}$ undertaken by the wind. {To leading order the
  pressure due to the wind is the base wind strength $W$ and the
  rate at which the wind performs work on the system corresponds
  to the product of pressure and volume flux. But, the integral of
  the volume flux over $\Delta t_{*}$ is simply the first term in}
\eqref{eq:dWa}, which corresponds to the last term in
\eqref{eq:dhdt_bnd} and \eqref{eq:h_bnd}:
\begin{equation}
  \Delta E_{w}=\underbrace{W}_{\text{pressure}}\underbrace{\frac{2V^{1/2}}{3}\Delta W_{*}^{3/2}}_{\text{volume}}.
\end{equation}
\noindent Equating $\eta\Delta E_{w}$ with $\Delta E_{p}$ implies
that
\begin{equation}
  \Delta W_{*}\sim \left(\frac{3}{4\eta}\right)^{2/3}\frac{1}{V^{1/3}W^{2/5}},
  \label{eq:energetics}
\end{equation}
\noindent {for $W\rightarrow\infty$, which is identical to}
\eqref{eq:dWa0} {when $\eta=1/2$. The fraction $\eta=1/2$
corresponds to the mixing efficiency of high-Rayleigh number
convection resulting from heating and cooling at the bottom and
top of a domain, respectively} \citep[see
e.g.][for the particular case of Rayleigh-B\'{e}nard
convection]{HugGjfm2013a}.
\section{Conclusions}

The transient response of a stratified interior to sudden changes
in the strength of the wind plays a significant role in
determining a system's robustness. In particular, the energy
required to homogenise a stratified interior makes existing
stability estimations based exclusively on a space's average
buoyancy overly conservative, especially in the case of large
opposing wind strengths.

Two- or multiple-layer stratifications can be produced by a
variety of localised sources of buoyancy, besides the point
sources considered here. Each would entail a slightly different
set of dynamical equations to account for the driven flow's
vertical variation in volume flux \citep[see e.g. line plumes
in][]{KayNjfm2004a}. However, our results are general in
suggesting that for moderate to large base wind strengths it is
the initial, steady-state, density profile that is dominant in
determining the system's robustness, rather than the particular
contribution of volume and heat to the upper layers during the
relatively rapid destruction of the stratification.

Further considerations are required to quantify the robustness of
a given state more precisely. For example, a low-level opening of
finite vertical extent is likely to erode the criterion for
stability that we have developed by permitting buoyancy to escape
from the interior before the interface reaches floor level
\citep[see e.g.][]{HunGjfm2005a}. On the other hand, the model
that we have used assumes that fluid entering the upper layer
mixes with the surrounding fluid {completely and
  instantaneously}. If relatively cool, dense air descended
through the space without mixing significantly with the upper
layer, the interface would not be lowered and the critical
increase in wind strength might be even larger than our
predictions suggest. {Whilst such effects are captured
  phenomenologically by values of $\eta$ in} \eqref{eq:energetics}
that are less than $1/2$, they warrant further attention.

\section*{Acknowledgements}

J.C. gratefully acknowledges an Imperial College Junior Research
Fellowship.

\addcontentsline{toc}{section}{References}
\bibliographystyle{jfm}
\bibliography{main}

\begin{thebibliography}{9}
\expandafter\ifx\csname natexlab\endcsname\relax\def\natexlab#1{#1}\fi

\bibitem[Coomaraswamy \& Caulfield(2011)]{CooIjfm2011a}
{\sc Coomaraswamy, I. \& Caulfield, C.} 2011 Time-dependent ventilation flows
  driven by opposing wind and buoyancy. {\em Journal of Fluid Mechanics\/} pp.
  33--59.

\bibitem[Gladstone \& Woods(2001)]{GlaCjfm2001a}
{\sc Gladstone, C. \& Woods, A.~W.} 2001 On buoyancy-driven natural ventilation
  of a room with a heated floor. {\em Journal of Fluid Mechanics\/} {\bf 441},
  293–314.

\bibitem[Hughes {\em et~al.\/}(2013)Hughes, Gayen \& Griffiths]{HugGjfm2013a}
{\sc Hughes, G.~O., Gayen, B. \& Griffiths, R.~W.} 2013 Available potential
  energy in {R}ayleigh-{B}\'{e}rnard convection. {\em Journal of Fluid
  Mechanics\/} {\bf 729}.

\bibitem[Hunt \& Linden(2005)]{HunGjfm2005a}
{\sc Hunt, G.~R. \& Linden, P.~F.} 2005 Displacement and mixing ventilation
  driven by opposing wind and buoyancy. {\em Journal of Fluid Mechanics\/} {\bf
  527}, 27--55.

\bibitem[Kaye \& Hunt(2004)]{KayNjfm2004a}
{\sc Kaye, N.~B. \& Hunt, G.~R.} 2004 Time-dependent flows in an emptying
  filling box. {\em Journal of Fluid Mechanics\/} {\bf 520}, 135--156.

\bibitem[Linden(1999)]{LinPafm1999a}
{\sc Linden, P.~F.} 1999 The fluid mechanics of natural ventilation. {\em
  Annual Review of Fluid Mechanics\/} {\bf 31}~(1), 201--238.

\bibitem[Linden {\em et~al.\/}(1990)Linden, Lane-Serff \& Smeed]{LinPjfm1990a}
{\sc Linden, P.~F., Lane-Serff, G.~F. \& Smeed, D.~A.} 1990 Emptying filling
  boxes: the fluid mechanics of natural ventilation. {\em Journal of Fluid
  Mechanics\/} {\bf 212}, 309--335.

\bibitem[Lishman \& Woods(2009)]{LisBbae2009a}
{\sc Lishman, B. \& Woods, A.~W.} 2009 On transitions in natural ventilation
  flow driven by changes in the wind. {\em Building and Environment\/} {\bf
  44}~(4), 666--673.

\bibitem[Yuan \& Glicksman(2008)]{YuaJbae2008a}
{\sc Yuan, J. \& Glicksman, L.~R.} 2008 Multiple steady states in combined
  buoyancy and wind driven natural ventilation: The conditions for multiple
  solutions and the critical point for initial conditions. {\em Building and
  Environment\/} {\bf 43}~(1), 62--69.

\end{thebibliography}

\end{document}